\documentclass[%
 aip,
 amsmath,amssymb,
 reprint,%
]{revtex4-1}

\usepackage{graphicx}
\usepackage{dcolumn}
\usepackage{bm}

\usepackage[utf8]{inputenc}
\usepackage[T1]{fontenc}
\usepackage{mathptmx}
\usepackage{etoolbox}
\usepackage{xr}
\usepackage{epsfig}

\makeatletter
\def\@email#1#2{%
 \endgroup
 \patchcmd{\titleblock@produce}
  {\frontmatter@RRAPformat}
  {\frontmatter@RRAPformat{\produce@RRAP{*#1\href{mailto:#2}{#2}}}\frontmatter@RRAPformat}
  {}{}
}%
\makeatother

\makeatletter
\newcommand*{\addFileDependency}[1]{
  \typeout{(#1)}
  \@addtofilelist{#1}
  \IfFileExists{#1}{}{\typeout{No file #1.}}
}
\makeatother

\newcommand*{\myexternaldocument}[1]{%
    \externaldocument{#1}%
    \addFileDependency{#1.tex}%
    \addFileDependency{#1.aux}%
}
\myexternaldocument{supplement.tex}

\begin{document}

\preprint{AIP/123-QED}

\title[]{Disentangling the magneto-optic Kerr effect of manganite epitaxial heterostructures}
\author{Jörg Schöpf}
\affiliation{University of Cologne, II Physics Institute, Cologne, Germany
}%
\author{Paul H. M. van Loosdrecht}
\affiliation{University of Cologne, II Physics Institute, Cologne, Germany
}%

\author{Ionela Lindfors-Vrejoiu}
\affiliation{University of Cologne, II Physics Institute, Cologne, Germany
}%

\date{\today}

\begin{abstract}
Magneto-optic Kerr effect can probe the process of magnetization reversal in ferromagnetic thin films and thus be used as an alternative to magnetometry. Kerr effect is wavelength-dependent and the Kerr rotation can reverse sign, vanishing at particular wavelengths. 
We investigate epitaxial heterostructures of ferromagnetic  manganite, La$_{0.7}$Sr$_{0.3}$Mn$_{0.9}$Ru$_{0.1}$O$_3$, by polar Kerr effect and magnetometry. The manganite layers are separated by or interfaced with a layer of nickelate, NdNiO$_3$. Kerr rotation hysteresis loops of trilayers, with two manganite layers of different thickness separated by a nickelate layer, have intriguing humplike features, when measured with light of 400 nm wavelength. By investigating additional reference samples we disentangle the contributions of the individual layers to the loops: we show that the humps originate from the opposite sense of the Kerr rotation of the two different ferromagnetic layers, combined with the additive behavior of the Kerr signal. 

\end{abstract}

\maketitle

\begin{quotation}

\end{quotation}

\section{\label{sec:level1}Introduction}
\begin{figure}
    \centering
    \includegraphics[scale=1]{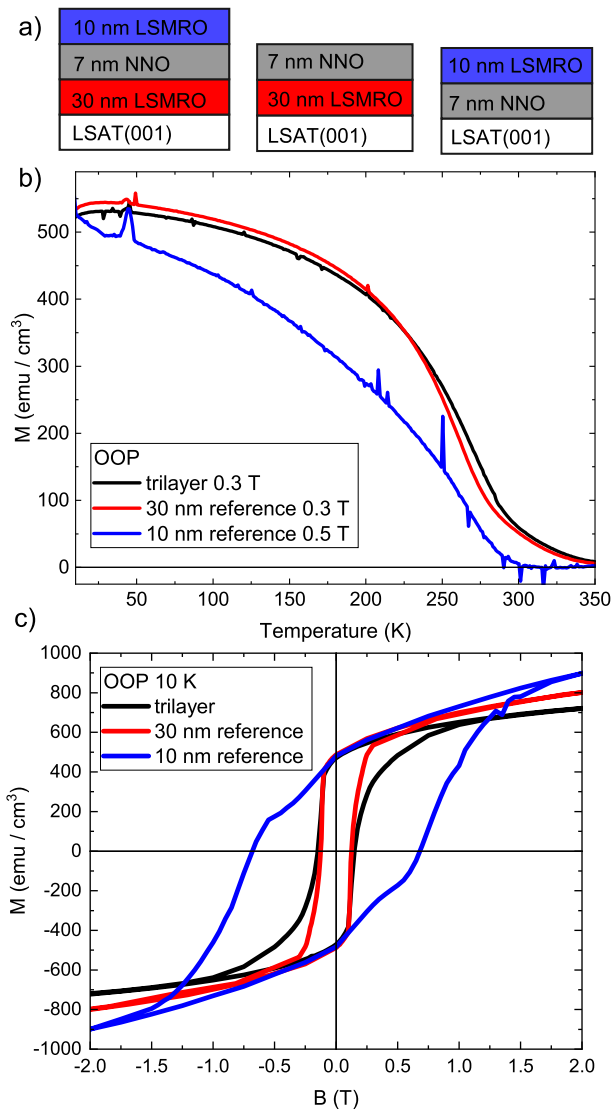}
    \caption{Sample description and SQUID magnetometry investigations of the heterostructures, with magnetic field applied perpendicular (OOP) to the layers. a) Sketches of the heterostructures: trilayer and the two reference samples corresponding to the bottom and top manganite layers of the trliayer. b) Temperature dependence of the  magnetization  of the trilayer (black), 30 nm thick reference (red) and 10 nm thick reference (blue), measured during field cooling in 0.3 or 0.5 Tesla (as indicated). c) Comparison of the magnetization hysteresis loops for the trilayer and the two reference samples at 10 K (color coding as above).}
    \label{fig:1}
\end{figure}

Magnetic properties of epitaxial ferromagnetic oxide heterostructures and multilayers are  governed by the interplay between magnetocrystalline anisotropy, interface and shape anisotropy, and by the magnetic interlayer coupling.\cite{Bhattacharya2014} In thin film systems these simultaneously affect the magnetic ordered phases as a function of temperature and applied magnetic field in an intricate way. Moreover, the evaluation of the magnetic interlayer coupling strength in such heterostructures is often problematic: one can assess it by performing minor magnetization hysteresis loops \cite{vanderHeijden1997} or first order reversal curves (FORC) magnetometry or MOKE studies \cite{Gilbert2021, Graefe2014}. SQUID magnetometry of magnetic thin films is, however, often affected by spurious contributions from the bulk substrates.\cite{Wysocki2022}  Magneto-optic Kerr effect (MOKE) measurements performed in reflection geometry can circumvent this difficulty. Concerning ferromagnetic oxide heterostructures, we investigated by SQUID magnetometry and MOKE the magnetic interlayer coupling between ferromagnetic SrRuO$_3$ epitaxial layers with perpendicular magnetic anisotropy separated by various spacer layers, such as SrIrO$_3$/SrZrO$_3$ \cite{Wysocki2018}, SrIrO$_3$ \cite{Wysocki2022} or LaNiO$_3$ \cite{Yang2021}. The type of spacer layer and its physical properties influence strongly the strength of the coupling. In particular, a LaNiO$_3$ spacer was a suitable choice for achieving strong ferromagnetic interlayer coupling. The coupling strength depended dramatically on the thickness of the LaNiO$_3$, as the spacer undergoes a metal-insulator transition at about 4 pseudocubic monolayers thickness. It is thus interesting to study the interlayer coupling when the spacer layer has a metal-insulator transition (MIT) as a function of temperature, as it is the case for NdNiO$_3$ (NNO), which as bulk has the MIT at about 200 K. \cite{Catalan2000, Catalano2018} For this purpose the ferromagnetic layers should have a Curie temperature (T$_C$) significantly larger than 200 K and therefore we chose  La$_{0.7}$Sr$_{0.3}$Mn$_{0.9}$Ru$_{0.1}$O$_3$ (LSMRO) layers (T$_C$ is about 300 K).\cite{Wang2007, Nakamura2018} Additionaly, epitaxial growth on (LaAlO$_3$)$_{0.3}$-(SrAl$_{0.5}$Ta$_{0.5}$O$_3$)$_{0.7}$ (LSAT(100)) substrates and Ru substitution renders LSMRO layer with perpendicular magnetic anisotropy. \cite{Nakamura2018} The perpendicular magnetic anisotropy is of importance as our aim is to study the heterostructures by polar MOKE measurements, in which the magnetic field is applied perpendicular to the sample surface. First we need to understand the polar MOKE response of a heterostructure with two ferromagnetic layers that have individual magneto-optic properties, which can be strongly wavelength-dependent. In particular, we have to understand how the Kerr rotation hysteresis loops relate to each of the layers of the heterostructure and to their magnetization hysteresis loops measured by magnetometry. This will enable the comparison of minor and full loops, which can be then employed to assess the strength of the interlayer coupling \cite{vanderHeijden1997}. \\
Therefore, here we focus on the polar MOKE response of trilayer samples with two LSMRO layers of different thickness separated by a NNO layer (see schematics in Fig.~\ref{fig:1}a). The Kerr loops of the trilayer measured with light of 400 nm wavelengh have an unusual shape with symmetric humplike features, not shown by the magnetization loops obtained by magnetometry. We compare the Kerr rotation loops of the trilayer with the loops of reference samples (see Fig.~\ref{fig:1}a) and measure the wavelength dependence of the Kerr rotation of the trilayer and the reference samples. We demonstrate that the apparently anomalous behavior of the MOKE loops of the trilayer is the simple result of the wavelength dependence of the magneto-optic properties and of the additivity of the
loops of the two different ferromagnetic layers.

\section{\label{sec:level1}Experimental results and discussion}
The layers of the heterostructures were made by pulsed-laser deposition (PLD). The as-received LSAT substrates of square shape and 4 mm size (CrysTec GmbH) were annealed at 1273 K for 2 hours in air prior to being used for PLD. Ceramic stoichiometric targets were used for laser ablation. The layers were grown with a laser fluence of 2.4 J/cm$^2$ and a laser pulse repetition rate of 5 Hz for LSMRO and 2 Hz for NNO, at 923 K in 0.133 mbar O$_2$ (LSMRO) and 0.3 mbar O$_2$ (NNO). After growth the samples were cooled down in 100 mbar O$_2$ with a rate of 10 K/min.
A homemade setup was used  for simultaneous polar MOKE and magnetotransport measurements. The MOKE measurements were done by employing a double modulation technique using an optical chopper and photoelastic modulator (PEM100, Hins Instruments). A Xe-lamp was used as a light source in conjunction with a Jobin Yvon monochromator.  Magnetotransport measurements were performed in a van der Pauw geometry and electrical contacts were done by gluing copper wires with silver paint on the corners of the samples. SQUID magnetometry was performed with a commercially available SQUID magnetometer (MPMS-XL, Quantum Design inc.).

\subsection{\label{sec:level2}SQUID magnetometry}
Figure ~\ref{fig:1} shows a summary of the SQUID magnetometry data (magnetization as a function of temperature and magnetic field, with the magnetic field applied perpendicular to the sample surface), for the three samples schematically depicted in Fig.~\ref{fig:1}a. The reference samples are used to mimic the top part and the bottom part of the sample of interest, the trilayer, and give us information about the magnetic properties and later of magneto-optic and magnetotransport properties of the analogue parts in the trilayer. The field cooled magnetization curves of Fig.~\ref{fig:1}b reveal a decrease of the Curie temperature of about 30 K for the 10 nm reference sample, for which the LSMRO layer is grown on top of the NNO layer. The 30 nm reference sample that mimics the bottom part of the trilayer, in which the 30 nm LSMRO layer grows directly on the LSAT substrate, has the same Curie temperature as for the trilayer. This already hints that the epitaxial growth of LSMRO on top of NNO influences drastically its magnetic properties, most likely via interfacial structural accommodations that affect the structure of the thin LSMRO layer and result in important physical property changes. This is further reflected by the magnetization hysteresis loops plotted in Fig.~\ref{fig:1}c: the  10 nm reference sample has a massively increased coercive field at 10 K, while the coercive field of the 30 nm reference sample matches well the coercive field of the trilayer. We summed up the magnetization loops of the two references and the comparison of this artificial loop to the loop of the trilayer is shown in the supplementary material (see \textbf{Fig. S1}): there is quite good agreement between the measured and the sum loops. We stress that the strongly slanted shape of the hysteresis loop of the 10 nm reference indicates a change of the magnetic anisotropy, the perpendicular direction (out-of-plane, OOP) is most likely not the easy axis for the 10 nm LSMRO layer grown on top of NNO. This explains also the shape of the hysteresis loop of the trilayer, where at higher fields a pronounced tail is observed before reaching saturation, corresponding to the gradual magnetization reversal in the top 10 nm LSMRO layer. We conclude that in the trilayer the two LSMRO have very different direction of the magnetization easy axis, with predominantly OOP for the bottom 30 nm LSMRO and a strongly in-plane tilted direction for the 10 nm LSMRO, the latter being induced by the growth on the NNO spacer. This conclusion is supported by the magnetic properties of bare 10 nm and 30 nm thick LSMRO layers grown directly on LSAT substrates (with no interfaces to NNO): both show clear OOP magnetic anisotropy, with square hysteresis loops with large remnant magnetization and low coercive fields (see supplementary material, \textbf{Fig. S2}).

\begin{figure}
    \centering
    \includegraphics[scale=0.9]{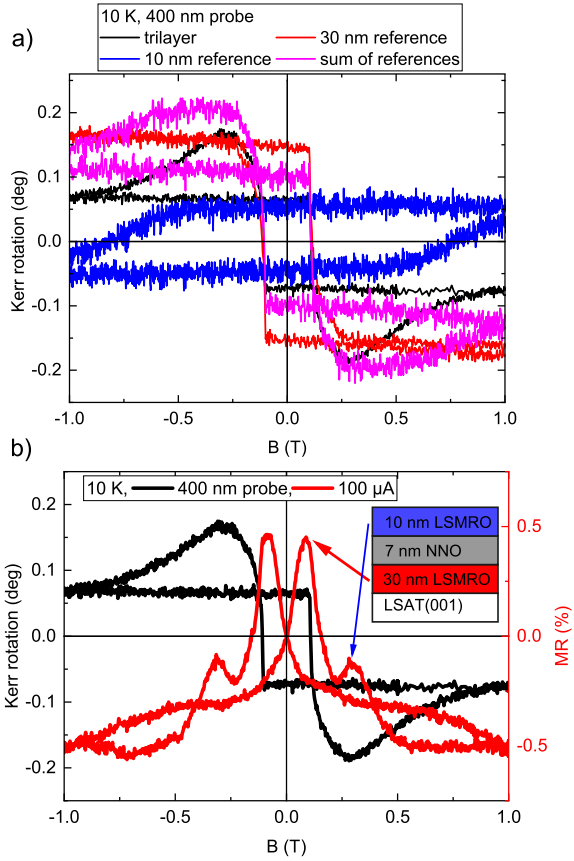}
    \caption{Polar MOKE investigations of the heterostructures and of the reference samples. a) Kerr rotation loops measured at 400 nm and at 10 K. The magenta loop is the sum of the loops of the loops of the two references.  b) Comparison of the Kerr loop  and of the linear magnetoresistance loop of the trilayer sample measured simultaneously at 10 K. In b) the arrows indicate that the linear magnetoresistance (MR) loop exhibits pronounced peaks, occurring at the two coercive fields of the different LSMRO layers of the trilayer, as shown in the inset schematics.}
    \label{fig:2}
\end{figure}

\begin{figure}
    \centering
    \includegraphics[width=8.2 cm]{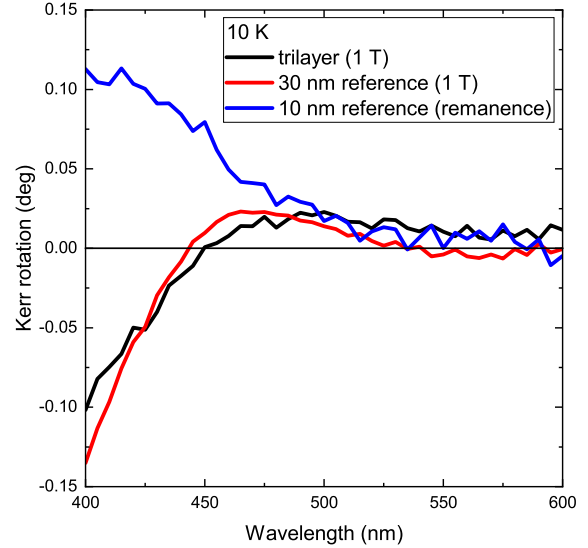}
    \caption{Polar Kerr rotation spectra of the heterostructures and of the reference samples at 10 K. Measurements were performed between 400 nm and 600 nm in 1 T and remanence (as indicated) with a stepsize of 5 nm.}
    \label{fig:3}
\end{figure}

\subsection{\label{sec:level3}Magneto-optic Kerr effect investigations}

The magneto-optic properties of the heterostructures were investigated by measuring the Kerr rotation hysteresis loops with a probe of 400 nm wavelength,  at 10 K (Fig.~\ref{fig:2}). One can note that both the trilayer and 30 nm reference have a negative sense of Kerr rotation in saturation in positive applied magnetic fields, while the 10 mn reference shows an (opposite) positive sense of the Kerr rotation. The Kerr rotation loop of the trilayer (plotted in black in Fig.~\ref{fig:2}a and b) has a peculiar behavior: starting in saturation at +1 T at about -70 mdeg, the Kerr rotation stays constant up to an applied field of -100 mT, where a sharp reversal of the magnetization of the 30 nm bottom layer, with lower coercivity, occurs (in agreement with the magnetization hysteresis loops in Fig. ~\ref{fig:1}). However, in contrast to the magnetization hysteresis loop of the trilayer, the Kerr rotation has a pronounced humplike feature,  observed from -290 mT to -1 T, after which  saturation occurs. Similar behavior was observed for the Kerr rotation loops of ferromagnetic SrRuO$_3$ films with inhomogeneous strain distribution.\cite{Bartram2020} For the Kerr loops of the trilayer, the apparent "hump" results from the opposite sense of the Kerr rotation of the 10 nm LSMRO top layer, with respect to the Kerr rotation of the 30 nm layer. The magnitude of total magnetization starts to increase as the 10 nm LSMRO top layer reverses its magnetization to be (mostly) parallel to magnetization of the 30 nm LSMRO bottom layer, but the total measured Kerr rotation of the trilayer decreases due to the opposite sense of the Kerr rotation of the different LSMRO layers in the trilayer. This can be readily proven by summing up the Kerr rotation loops of the 30 nm reference and 10 nm reference and noting that $\theta_{\mathrm{total}} \approx \theta_{\mathrm{30 nm\: reference}} + \theta_{\mathrm{10 nm\: reference}}$, where $\theta$ denotes the measured Kerr rotation. \cite{phdthesis_Hemrle} The summation of the Kerr rotation hysteresis loops qualitatively reproduces the measured Kerr rotation loop of the trilayer (see the  magenta loop in Fig.~\ref{fig:2}a), although not perfectly. Slight differences are to be expected, due to multiple reasons: for the two reference samples the overall thickness of the heterostructure is different than that of the trilayer; for the 10 nm reference the NNO layer is grown on top of the LSAT substrate directly, not on top of the 30 nm LSMRO, as in case of the trilayer and thus the NNO spacer layer of the trilayer and the NNO bottom layer of the 10 nm reference may have different properties.   
To verify the difference in the sign of the Kerr rotation at 400 nm of the two reference samples, Kerr rotation spectra were measured between 400 nm and 600 nm (see Fig.~\ref{fig:3}) for the trilayer and reference samples at 10 K. For the trilayer and the 30 nm reference, a sign change of the Kerr rotation occurs around 450 nm from negative at lower wavelengths to positive at higher wavelengths. This spectral behavior is consistent with previous studies of La$_{2/3}$Sr$_{1/3}$MnO$_3$ on SrTiO$_3$ and arises from magneto-optic active transitions: an intra-3\textit{d} Mn crystal field transition around 460 nm and a charge transfer from O 2\textit{p} to Mn 3\textit{d} \textit{e}$_{\mathrm{g}}$ transition at 345 nm. The Kerr rotation of the 10 nm reference is positive at all wavelengths in the measurement range, which is consistent with the behavior of the measured Kerr rotation loops and indicating a change in the electronic structure of the 10 nm LSMRO film when grown epitaxially on NNO. The origin of this change in physical properties is most likely structural and requires further investigations. Spectra of 10 nm and 30 nm thick bare LSMRO films grown directly on LSAT and without NNO layers show similar behavior as the 30 nm reference and the trilayer, with the change of sign (see \textbf{Fig. S3} in the supplementary material).  We note that analysis of the ellipticity loops, measured simultaneously with the Kerr rotation loops (see \textbf{Fig. S4} in the supplementary material), gives us further confidence in our conclusion concerning the origin of the hump features.

\section{\label{sec:level1}Summary}

In summary,  we disentangled the behavior of the Kerr rotation loops of epitaxial trilyers with two ferromagnetic manganite layers separated by a nickelate spacer: the Kerr rotation loops measured at 400 nm wavelength showed intriguing humplike features in the magnetic field region where the top ferromagnetic layer reverses its magnetization. In order to unravel the origin of the humps, we compared the Kerr rotation loops with magnetization loops measured by SQUID magnetometry, and we saw that the hump of the Kerr loops corresponds to the tail of the SQUID loops before reaching saturation in high fields. We made reference samples that correspond to the trilayer lower part (with the 30 nm thick LSMRO and a top NNO) and upper part (with a NNO and a 10 nm thick LSMRO), and measured the SQUID and Kerr rotation loops of these two parts independently. This enabled us to probe the magnetic properties of the independent ferromagnetic layers of the trilayer and then understand the magnetization reversal in the trilayer: upon applying perpendicular magnetic field, the 30 nm LSMRO bottom layer that has predominantly OOP magnetic anisotropy and lower coercivity reverses first its magnetization ; the top 10 nm LSMRO, which has more towards in-plane tilted magnetic ansiotropy and enhanced coercivity, reverses the magentization at much larger fields and its magnetization reversal results into the tail of the SQUID loop and into the hump of the Kerr loop.  The hump is also the result of the different sign of the Kerr rotation of the two manganite layers at 400 nm wavelength: it is negative for the bottom layer and positive for the top layer. The change of sign and the magnetic anisotropy difference between the two manganite layers of the trilayer are consequences of the epitaxial growth of the 10 nm LSMRO on the nickelate spacer and require further structural and electronic structure investigations. Our findings stress how important the interfacial effects can be for the effective magnetic anisotropy and for the magneto-optic properties of epitaxial ferromagnetic oxide heterostructures.\\

See the supplementary material for the magnetization and Kerr rotation loops and Kerr rotation spectra of bare 10 nm thick and 30 nm LSMRO films grown directly on LSAT substrates, and for a discussion on the Kerr ellipticity loops.

\section*{Authors declarations}
Conflict of Interest\\
The authors have no conflicts to disclose.\\

Author Contributions\\
\textbf{J. Schöpf} performed the physical property measurements and data analyses and wrote the paper. \textbf{P. van Loosdrecht} participated in the MOKE data analyses and interpretation, and contributed to the paper writing.  \textbf{I. Lindfors-Vrejoiu} conceived the research, made the samples, and wrote the paper.  \\

\begin{acknowledgements}
We wish to acknowledge the financial support from the German Research Foundation (DFG) within SFB1238, project A01 (Project No. 277146847).
\end{acknowledgements}

\section*{Data Availability Statement}
The data that support the findings of this study are available from the corresponding authors upon reasonable request.

\section*{References}
\nocite{*}
\bibliography{LSMRONNORef}

\end{document}